# Galaxies, Human Eyes and Artificial Neural Networks

O. Lahav[1], A. Naim[1], R. J. Buta[2], H. G. Corwin[3], G. de Vaucouleurs[4], A. Dressler[5], J. P. Huchra[6], S. van den Bergh[7], S. Raychaudhury[6], L. Sodré Jr.[8] & M. C. Storrie-Lombardi[1]

[1] Institute of Astronomy, Madingley Rd., Cambridge, CB3 0HA, UK

[2] Dept. of Physics and Astronomy, University of Alabama, Tuscaloosa, AL, 35487-0324 USA

[3] California Institute of Technology, IPAC M/S 100, Pasadena, CA, 91125, USA

[4] Dept. of Astronomy, RLM 15.308, University of Texas, Austin, TX 78712-1083, USA

[5] Carnegie Observatories, 813 Santa Barbara street, Pasadena, CA, 91101-1292, USA

[6] Harvard-Smithsonian Center for Astrophysics, 60 Garden street, Cambridge, MA, 02138, USA

[7] Dominion Astrophysical Observatory, 5071 W. Saanich Rd., Victoria, BC, V8X 4M6, Canada

[8] Instituto Astronômico e Geofísico da Universidade de São Paulo, CP9638, 01065, São Paulo, Brazil

*Quantitative morphological classification of galaxies is important for understanding the origin of type frequency and correlations with environment. But galaxy morphological classification is still mainly done visually by dedicated individuals, in the spirit of Hubble's original scheme, and its modifications. The rapid increase in data on galaxy images*

*at low and high redshift calls for re-examination of the classification schemes and for new automatic methods. Here we show results from the first systematic comparison of the dispersion among human experts classifying a uniformly selected sample of over 800 digitised galaxy images. These galaxy images were then classified by six of the authors independently. The human classifications are compared with each other, and with an automatic classification by Artificial Neural Networks (ANN). It is shown that the ANNs can replicate the classification by a human expert to the same degree of agreement as that between two human experts.*

Hubble (1) suggested a classification scheme for galaxies which consists of a sequence starting from elliptical galaxies (E), through lenticular (S0), to spiral galaxies (S), and a parallel branch of spirals with a barred component, leading to the so called 'tuning fork' Hubble diagram. This scheme has been extended by astronomers over the years (2–5), to incorporate features such as the strength of the spiral arms, yielding multi-dimensional classification (3,5). It is remarkable that these somewhat subjective classification labels for galaxies (as seen projected on the sky) correlate well with physical properties such as colour, dynamical properties (e.g. rotation curves and stellar velocity dispersions) and the mass in neutral hydrogen (6). However, one would like eventually to devise a scheme of classification, which can be related to the physical processes of galaxy formation. While there have been in recent years significant advances in observational techniques (e.g. telescopes, detectors and reduction algorithms) as well as in theoretical modelling (e.g. N-body and hydrodynamics simulations), galaxy classification remains a subjective area.

Quantifying galaxy morphology is important for various reasons. First, it provides important clues to the origin of galaxies and their formation processes. For example, ellipticals and lenticular galaxies comprise only $\sim 20\%$ of the galaxies, and there is a striking density-morphology relation (1,7), indicating that elliptical galaxies mainly re-side in high-density regions. Understanding the origin of the type frequency and the density-morphology relation is clearly of fundamental importance. But quantifying these properties requires reliable classification schemes. Second, galaxies can also be used e.g. to measure redshift-independent distances by methods such as the luminosity-rotation velocity relation for spirals (8) and the diameter-velocity dispersion for ellipticals (9). Clearly any observational programme requires an *a priori* target list of objects for photometric or spectrographic measurements. Therefore galaxy classification is important for both practical reasons of producing large catalogues for statistical and observational programs, as well as for establishing some underlying physics (in analogy with the Hertzsprung-Russell diagram for stars). Moreover, understanding the morphology of galaxies at low redshift is crucial for any meaningful comparison with galaxy images obtained with the Hubble Space Telescope at higher redshift ($z \sim 0.4$). Most of our current knowledge of galaxy morphology is based on the pioneering work of several dedicated observers who classified thousands of galaxies and catalogued them (2, 10, 11). However, facilities such as the Cambridge Automated Plate Measuring (APM) machine and the Sloan digital sky survey yield millions of galaxies. Classifying very large data sets is obviously beyond the capability of a single person. Therefore, the galaxy classification problem calls for new approaches (12 − 16).

As a first step towards finding an automated method of galaxy classification, we compiled a well-defined sample of galaxy images. The galaxies were chosen from the APM Equatorial Catalogue of galaxies (17). This sample was compiled from IIIaJ (broad blue-green band) plates taken with the UK Schmidt telescope at Siding Spring, Australia, covering most of the sky between declinations $-17°.5 < \delta < 2°.5$, and Galactic latitudes $b \geq 20°$. We chose a subsample of galaxies with major diameter (at an isophotal level of 24.5 magnitudes per arcsec$^2$) $D \geq 1.2$ arcmin on 75 plates, after eliminating galaxies that had severe contamination from overlapping stellar or galaxy images ($< 10\%$). This sample of 831 galaxies was scanned in raster mode at a resolution of 1 arcsec by the APM (although the actual resolution of the Schmidt plates was more like 2 arcsec due to observing conditions). The digitised images (most of them of $256 \times 256$ pixels) were printed at full resolution.

The same galaxy images were then classified by six of the authors (RB, HC, GV, AD, JH and vdB) according to the Revised Hubble $T$-type (numerical stage) system (see RC3 (10) and Table 1 for details), or converted to it. Although the $T$-type is only a one-dimensional parameter (extending from $T = -6$ to $T = +11$) in a three-dimensional scheme (10), it is commonly used and is convenient for computer algorithms compared with other, more descriptive schemes. While five of the authors classified the images on laser-printed hard copies, vdB examined them on a computer screen. His classification was done according to the DDO system (5), which was then converted to the $T$-type (10).

The motivation for performing a comparison between different experts is two-fold. (i) To study systematically the degree of agreement and reproducibility between observers. (ii) To use the human classifications as 'training sets' for Artificial Neural Networks and other automatic classifiers. To our knowledge, this study is the first systematic comparison based on a uniform sample of galaxy images, and presented to a large number of experts, from different 'schools of thought'.

Figure 1 shows the digitised images of four galaxies in our sample. We also give the classification assigned to these galaxies by the RC3 catalogue (10) (ignoring the quoted uncertainty in their $T$-type), and by the six authors who independently classified the galaxies. One of these galaxies got exactly the same classification by all six observers, whereas there was no such clear agreement on the other three galaxies. Statistically, all six authors agreed on the exact $T$-type for only 8 galaxies out of the 831 (i.e. less than 1 %). Agreement between pairs of observers in excess of 80 % is obtained only to within 2 types. GV and vdB, who classified galaxies over many more years than the others, were rather conservative and did not classify about a third of the galaxy images which are saturated or of low quality. The other observers were more liberal and classified almost all the galaxies (see the second row in Table 3). On the whole, there is indeed a reasonable consistency in the way people classify galaxies, but the scatter is significant.

To better quantify the degree of agreement between observers we calculated for each pair of observers $a$ and $b$ the variance

$$\sigma_{ab}^2 = \frac{1}{N_{ab}} \sum_{gal} (T_a - T_b)^2 , \qquad (1)$$

taking into account only those $N_{ab}$ galaxies for which both observers gave a classification.

Table 2 shows the rms dispersion $\sigma_{ab}$ between all pairs of observers and also with the RC3 sample (10). Clearly the dispersion between RC3 and any of the observers (2.2 $T$-units on average) is larger than between any two observers who looked at the *same* APM images (1.8 $T$-units on average). We note that the subset of 600 RC3 galaxies in the sample has a median diameter of 1.7 arcmin, compared with the median 1.5 arcmin of the entire sample of 831 galaxies, and the images were on different plate materials. This illustrates the fact that any classification depends on the colour, size and quality of the images used, i.e. there is no 'universal' classification.

Another interesting result is that observers who belong to the same 'school' agree better with each other than with others. For example, the dispersion between GV and HC is only 1.5 and between HC and RB only 1.3 units. This indicates that systematic 'training' can reduce the scatter between two human experts. We also notice a weak trend for better agreement in classification for galaxies which are large (the rms dispersion between experts drops by about 10 % from 1.2 arcmin to 2 arcmin galaxies), but there is no obvious trend as a function of eccentricity. Detailed analysis of this comparison will appear elsewhere (18). We also intend to evaluate the 'internal scatter' $\sigma_a$ (i.e. reproducibility) of each observer, when classifying again the same data set or a set with lower resolution. As a crude estimate, if we assume that $\sigma_{ab}^2 = \sigma_a^2 + \sigma_b^2$ we find for the different observers $\sigma_a$ in the range 1.0–1.5 . It is worth emphasising that the plate material used here suffers from problems of saturation, and the digitization of the images (although at pixel size of 1 arcsec) may have degraded the agreement between observers. Nevertheless, the plate material we have used is typical in many extragalactic studies.

Having established the degree of agreement between human experts, the challenge is to design a computer algorithm which will reproduce classification to the same degree a student or a colleague of the human expert can do it. Such an automated procedure usually involves two steps: (i) feature extraction from the digitised image, e.g. the galaxy profile, the extent of spiral arms, the colour of the galaxy, or an efficient compression of the image pixels into a smaller number of coefficients (e.g. Fourier or Principal Component Analysis). (ii) A classification procedure, in which a computer 'learns' from a 'training set' for which a human expert provided his or her classification.

Artificial Neural Networks (ANN), originally suggested as simplified models of the human brain, are computer algorithms which provide a convenient general-purpose framework for classification (19), including astronomical applications (20, 21). One commonly used ANN configuration consists of nodes arranged in a series of layers and utilizes the Backpropagation minimization algorithm (22). In Figure 2 we show a configuration in which the galaxy parameters are fed into the input layer, and the $T$-type classification appears as a single continuous output. The 'hidden layer' allows non-linear boundaries in a complicated parameter space. In the 'training' phase the free parameters of the network ('weights') are determined by least-square minimization of the difference between the calculated and true (i.e the expert's) type. Other network configurations are possible, including multiple output nodes which can provide Bayesian *a posteriori* probabilities for each class (14, 23).

Pilot studies (14, 23) utilized ANNs to classify about 5200 galaxies from the ESO-LV catalogue (11), using 13 parameters, illustrating that galaxies can be classified automati-cally, with rms dispersion of 2.1 $T$-units between the ANN and the experts (Lauberts & Valentijn). However, due to the lack of quantitative measure of dispersion among human experts for comparison, it was difficult for us to judge if the achieved success-rate was satisfactory. We have now applied (24) the same technique to our new APM sample, after extracting significant features (ellipticity, surface brightness, luminosity profile parameters, arms to disk ratio, concentration indices and arms parameters) from the images. We then trained the ANN on the $T$ system (as in the network shown in Figure 2), feeding as input 13 parameters and allowing 5 nodes at the 'hidden layer', i.e. a 13:5:1 configuration (other network configurations have also been used, e.g. 13:13:1, yielding similar results). For each of the six individual expert classifications, the ANN was trained on $\frac{3}{4}$ of the sample and tested on the remaining $\frac{1}{4}$. As the ANN minimization begins with a set of random weights, we repeated the training and testing 10 times, with different initial weights (typical 'internal' scatter when nets with different initial random weights are used is about 0.1-0.3 units). The same process was repeated for the other three quarters of each set, resulting in 40 runs for each expert classification.

The rms dispersion (cf. eq. 1) between the ANN and each expert is given in Table 3, quantifying to what extent the human classification can be *reproduced* by the computer algorithm. The rms dispersion varies between 1.9 to 2.3 $T$-units over the six experts. This relatively small variation from one expert to another is not too surprising. The number of galaxies classified by each of the experts was different (see the second row of Table 3), with bias towards face-on galaxies in some cases. A large rms dispersion may not necessarily reflect inconsistency in the expert's own classification, but rather a poorer fit between the human classification, the chosen parameters and the model (i.e. the ANN). A better agreement, 1.8 $T$-units, is achieved when the ANN is trained and tested on the mean type as deduced from all available expert classifications (after removing few outliers). Comparison of Tables 2 and 3 show remarkable similarity in the dispersion between two human experts and that between ANN and experts. In other words, our results indicate that the ANNs can replicate the expert's classification of the APM sample as well as other colleagues or students of the expert.

Figure 3 shows an example of the ANN vs. mean expert classification for 207 galaxies, after training on the remaining 624 galaxies in the sample (again averaging results from 10 runs with different initial random weights). The solid circles indicate galaxies larger than the median diameter of 1.5 arcmin, while the open circles indicate smaller galaxies. As all galaxies in our sample are larger than 1.2 arcmin there is no obvious trend for worse classification for smaller diameters, which is expected for much smaller galaxies. There is also no dramatic trend with ellipticity. Of the 831 galaxies classified by the ANN by the above procedure, 9 % deviate from the 'true' mean answer by at least 3 types. Most of them are very late types and irregulars ($T > 7$).

To summarize, we have presented a first systematic comparison of galaxy classification by six observers for a new large data set of galaxy images. The comparison indicates that while the $T$-system is convenient, the scatter between observers is non-negligible. Caution is called for in assuming a 'universal' frequency type distribution in comparison with models and with high-redshift galaxies. The observed frequency distribution clearly depends on the plate material and on the human expert. It is shown that the ANNs can replicate the classification by a human expert to the same degree of agreement as that between two human experts, about 1.8 $T$-units. Future work will focus on 'supervised' ANNs to preserve human experience in multi-dimensional classification $(3, 5)$, and on 'unsupervised' algorithms (e.g. by generalizing Principal Component Analysis to non-linear mapping) to define a 'new physical Hubble sequence' without any prior human classification.

**Acknowledgements**. We thank the UKST unit of the Royal Observatory of Edinburgh for the plate material, the APM group at RGO Cambridge for scanning support, M.

Irwin and D. Lynden-Bell for helpful discussions. We are also grateful to the anonymous referee for helpful comments.

18 Naim, A. et al., *Mon. Not. R. astr. Soc.*, submitted (1994).

19 Hertz, J., Krogh, A., & Palmer, R.G., *Introduction to the Theory of Neural Computation* (Addison-Wesley, Redwood city, California, 1991).

20 Odewahn, S.C., Stockwell, E.B., Pennington, R.L., Humphreys, R.M. & Zumach, W. A., *Astr. J.*, **103**, 318 (1991).

21 Serra-Ricart, M., Calbet, X., Garrido, L. & Gaitan, V., *Astr. J.*, **106**, 1685 (1993).

22 Rumelhart, D.E., Hinton, G.E. & Williams, R.J., *Nature*, **323**, 533 (1986).

23 Lahav, O., Naim, A., Sodré Jr., L. & Storrie-Lombardi, M.C., in preparation

24 Naim, A., Lahav, O., Sodré Jr., L. & Storrie-Lombardi, M.C., in preparation

**Table 1** The $T$-type system of galaxy classification (from RC3 (10)).

| $cE$ | $E0$ | $E^+$ | $S0^-$ | $S0^o$ | $S0^+$ | $S0/a$ | $Sa$ | $Sab$ | $Sb$ | $Sbc$ | $Sc$ | $Scd$ | $Sd$ | $Sdm$ | $Sm$ | $Im$ | $cI$ | $I0$ | $Pec$ |
|---|---|---|---|---|---|---|---|---|---|---|---|---|---|---|---|---|---|---|---|
| −6 | −5 | −4 | −3 | −2 | −1 | 0 | 1 | 2 | 3 | 4 | 5 | 6 | 7 | 8 | 9 | 10 | 11 | 90 | 99 |

**Table 2** The rms dispersion in $T$-type classification between pairs of observers

|     | RB  | HC  | GV  | AD  | JH  | vdB |
|-----|-----|-----|-----|-----|-----|-----|
| RC3 | 2.2 | 2.1 | 1.8 | 2.3 | 2.2 | 2.4 |
| RB  |     | 1.3 | 1.6 | 1.7 | 1.8 | 1.7 |
| HC  |     |     | 1.5 | 1.8 | 1.9 | 1.9 |
| GV  |     |     |     | 1.7 | 1.8 | 1.9 |
| AD  |     |     |     |     | 2.1 | 1.8 |
| JH  |     |     |     |     |     | 2.0 |

**Table 3** The rms dispersion in $T$-type classification between the ANN and human experts. The ANN was trained and tested on individual observers and their mean classification. The second row gives the total number of galaxies classified by each expert.

|           | RB  | HC  | GV  | AD  | JH  | vdB | Mean |
|-----------|-----|-----|-----|-----|-----|-----|------|
| ANN       | 1.9 | 2.0 | 2.2 | 1.9 | 2.3 | 2.2 | 1.8  |
| $N_{gal}$ | 764 | 812 | 473 | 814 | 824 | 549 | 831  |

**Figure 1** Four APM galaxy images and their classification by six of the authors and RC3. The $T$-type classification of NGC2811 by (RC3, RB, HC, GV, AD, JH, vdB) is (1, 1, 1, 1, 1, 1, 1), of NGC3200 (4.5, 5, 5, 4, 5, 4, 3), of NGC4902 (3, 3, 4, 3, 3, 5, 3) and of NGC3962 (-5, -3, 0, -5, -3, -1, -5).

**Figure 2** A schematic diagram of an Artificial Neural Network for classifying galaxies. In this configuration the galaxy parameters are fed into the input layer, and the $T$-type classification appears as a single continuous output. The network is trained according to classification by a human expert. The 'hidden layer' allows non-linear boundaries in a complicated parameter space.

**Figure 3** The ANN vs. mean expert $T$-type classification for 207 galaxies. The ANN was trained on the remaining 624 galaxies in the sample (with results averaged over 10 runs with different initial random weights). The solid circles indicate galaxies larger than the median diameter of 1.5 arcmin, while the open circles indicate smaller galaxies.